\documentclass[fleqn,10pt]{wlscirep}
\usepackage{graphicx,subfigure,xcolor,soul}
\usepackage{bm}

\title{Resonance interaction energy between two entangled atoms in a photonic bandgap environment}

\author[1]{Valentina Notararigo}
\author[2,3,*]{Roberto Passante}
\author[2,3]{Lucia Rizzuto}
\affil[1]{Department of Physics and Astronomy, University College London, Gower Street, London WC1E 6BT, UK}
\affil[2]{Dipartimento di Fisica e Chimica, Universit\`{a} degli Studi di Palermo, Via Archirafi 36, I-90123 Palermo, Italy}
\affil[3]{INFN, Laboratori Nazionali del Sud, I-95123 Catania, Italy}

\affil[*]{roberto.passante@unipa.it}

\begin{abstract}
We consider the resonance interaction energy between two identical entangled atoms, where one is in the excited state and the other in the ground state. They interact with the quantum electromagnetic field in the vacuum state and are placed in a photonic-bandgap environment with a dispersion relation quadratic near the gap edge and linear for low frequencies, while the atomic transition frequency is assumed to be inside the photonic gap and near its lower edge. This problem is strictly related to the coherent resonant energy transfer between atoms in external environments.
The analysis involves both an isotropic three-dimensional model and the one-dimensional case. The resonance interaction asymptotically decays faster with distance compared to the free-space case, specifically as $1/r^2$ compared to the $1/r$ free-space dependence in the three-dimensional case, and as $1/r$ compared to the oscillatory dependence in free space for the one-dimensional case.
Nonetheless, the interaction energy remains significant and much stronger than dispersion interactions between atoms. On the other hand, spontaneous emission is strongly suppressed by the environment and the correlated state is thus preserved by the spontaneous-decay decoherence effects.  We conclude that our configuration is suitable for observing the elusive quantum resonance interaction between entangled atoms.
\end{abstract}

\def\br{{\bf r}}
\def\wk{\omega_k}
\def\wa{\omega_a}
\def\akj{a_{\bk j}}
\def\akjd{a_{\bk j}^\dagger}

\def\ekj{\hat{e}_{\bk j}}

\def\bk{{\bf k}}

\def\k0{k_0}

\begin{document}

\flushbottom
\maketitle
\thispagestyle{empty}

\section*{Introduction}

The resonance interaction energy and the energy transfer between two identical atoms, one in the ground state and the other in the excited state, coupled with the electromagnetic field in the vacuum state, have recently received great attention in the literature in different research fields \cite{Salam08,ElGanainy-John13,Incardone-Fukuta14,Andrews-Ford13}. For example, coherent energy transfer is supposed to play an important role in the initial stages of photosynthesis \cite{Scholes03,OlayaCastro-Nazir12}. From a quantum-electrodynamical point of view, the resonance interaction is due to the exchange of a real or virtual photon between the two atoms.

If the two atoms are in a factorized state, their radiation-mediated interaction energy is a fourth-order process in the electric charge
\cite{Compagno-Passante-Persico95,Craig-Thirunamachandran98},
and controversial results exist in the literature concerning with the presence of spatially oscillating terms, when one atom is in an excited state
\cite{Donaire-Guerot15,Berman15,Barcellona-Passante16,Milonni-Rafsanjani15}. However, the interaction is a second-order distance-dependent energy shift when two identical atoms are prepared in a symmetric or antisymmetric entangled state
\cite{Salam08,Craig-Thirunamachandran98,Salam10}.
The resonance interaction is particularly relevant because it is much more intense than dispersion interactions (van der Waals/Casimir-Polder, both for ground-state and excited atoms) and Casimir-Polder interactions between a ground-state atom and a mirror, and comparable to the atom-surface interaction for excited atoms \cite{Salam08,Messina-Passante08,Haakh-Intravaia09}. The Casimir-Polder interaction between an excited atom and a surface has recently been measured directly
\cite{Intravaia-Henkel11,Wilson-Bushev03}.
From a physical point of view, the resonance interaction is a second-order effect because the correlated state has non-vanishing dipole-dipole correlations, while, for dispersion interactions, correlated dipole moments must be induced by vacuum field fluctuations \cite{Passante-Persico03,Power-Thirunamachandran93,Rizzuto-Passante07,Passante-Persico06,Bartolo-Passante12}. Moreover, the resonance interaction is of very long range; in free space and at very large distances (far zone), it scales as $r^{-1}$ in the three dimensional case, with spatial oscillations, while it is a purely oscillating function for a one-dimensional system \cite{Incardone-Fukuta14,Salam08}.

However, the resonance interaction requires the system be prepared in an entangled state and that the coherence of this state be preserved over a sufficiently long time: this makes it extremely difficult to observe. In fact, this quantum correlated state is very fragile, and the coherent superposition can be easily destroyed by the influence of the environment and by spontaneous emission. As far as we know, there has not yet been any direct observation of this interaction and the resulting interatomic force.
A somehow related effect, the coherent dipolar-induced exchange of excitation (F\"{o}rster resonance), has been recently observed for a system of two Rydberg atoms in the short-distance regime \cite{Ravets-Labuhn14}.
It is thus a fundamental issue to investigate setups that could allow observation and measurement of the resonance force, an elusive quantum coherent phenomenon, in particular in the long-distance regime.
We propose that an appropriate environment around quantum emitters (atoms, molecules or quantum dots), could preserve the correlated atomic state for a sufficiently long time and could therefore be exploited to observe and measure the resonance force between two quantum emitters for the first time.

The observation of the resonance force, in particular in the retarded large-distance regime, would be an important confirmation of an effect due to the quantum coherence of the system. It could also be relevant in different physical processes and in other fields, such as biology. For example, the possible fundamental role of resonant interactions in long-distance biomolecular recognition has been recently argued \cite{Preto-Pettini15}.
An analogous interaction, mediated by the exchange of an electron in the continuum band of the wire, has been shown for two impurities (adatoms) embedded in a semiconductor quantum wire \cite{Tanaka-Passante13,Yang-Yang16}.

In a recent paper, the interatomic resonance force has been shown to be strongly affected by the presence of a structured environment such as a photonic crystal that changes the photon dispersion relation and density of states, and determines a photon bandgap \cite{Incardone-Fukuta14}. In the configuration under scrutiny in that paper the atomic transition frequency is outside the photon bandgap edge, but very close to it. In this case, for a one-dimensional photonic crystal and for an isotropic three-dimensional crystal, the force can be significantly enhanced up to a factor $10^3$. However, the spontaneous emission rate increases too, making the entangled state of the two atoms even more fragile. As a result, on the one hand the resonance force can be more intense than in free space, but on the other hand, the correlated state decays quicker, making observation of the resonance force more difficult.

In this paper, we focus on two identical atoms in a generic environment with a photonic bandgap, whose transition frequency is now inside the bandgap and in the proximity of its lower edge. The model used relies on a dispersion relation which is quadratic for the photons near the edges of the band (effective mass approximation) and linear for low frequencies. The structured environment is assumed to be periodic in one dimension and homogeneous in the other two.
Importantly, inside the bandgap the photon density of states is strongly reduced and therefore the spontaneous decay is inhibited (see Refs. \cite{John-Wang90,Angelakis-Knight04,John-Wang91} for the case of a photonic crystal).
Inhibited decay rates of spontaneous emission have been recently observed for a quantum dot in a three-dimensional photonic crystal \cite{Leistkow-Mosk11}. Within photonic crystals a suppression of the spontaneous decay rate of more than a factor $10$ has been obtained in the optical region \cite{Noda-Fujita07,Jorgensen-Galusha11}.
Several different one-dimensional engineered environments exhibit a photonic bandgap, for example an array of coupled optical cavities, nanophotonic waveguides, photonic crystals, coupled transmission line resonators (for a review, see Ref. \cite{Lodahl-Mahmoodian15}).
Ultracold atoms in one-dimensional optical lattices have been also used to realize periodic structures \cite{Modugno2016,Xue2016}.
The relevance of the configuration considered here is that the spontaneous emission of the atoms is suppressed, and this allows the correlated state to be maintained for a time long enough to observe the resonance interaction.
In addition our result shows that if on the one hand the resonance interaction decays faster with the distance, compared to the free-space case in the long-distance regime, on the other hand it remains significant and it is also several orders of magnitude larger than the dispersion interaction between two atoms in a factorized state.
This is due to the fact that also off-resonance modes of the field contribute to the resonance force. In particular, low-frequency modes with a linear dispersion law are relevant for its asymptotic behavior, similarly to dispersion interactions, while only near-resonance modes (suppressed by the environment) are responsible of the spontaneous decay.
In a three-dimensional isotropic system, we find that the resonance interaction energy at large distances scales as $r^{-2}$ rather than the $r^{-1}$ free-space behavior.
Instead in the one-dimensional case and in the asymptotic limit it decreases as $r^{-1}$ rather than the pure oscillation of the free-space case.
We have also verified this result by a numerical evaluation using the exact dispersion relation of the photonic crystal.
For typical parameters of quantum emitters and photonic crystals the strength of the interaction is not significantly reduced at relevant distances in both these configurations.
All this indicates that our configuration can be an appropriate experimental setup for observing the resonance interaction.
Controversial results have been obtained in the literature regarding the suppression of the dipole-dipole interaction and energy transfer in bandgap materials, when the frequency of the two atoms is inside the forbidden band \cite{Kurizki-Genack88,Zheng-Kobayashi96,Kurizki-Kofman96,Kurizki90}.

Other recent results on the dipole-dipole resonant interaction between two atoms in photonic crystals waveguides indicate an exponential decay when the atomic frequency is within the gap of a photonic crystal waveguide \cite{Shahmoon-Kurizki13,Douglas-Habibian15,Shahmoon-Grisins16} or in photonic crystals \cite{John-Wang91}. Our results, instead, show a power-law decay of the resonance interaction energy. We argue that the difference in the scaling with the distance is ultimately related to the different structure of the photonic modes in the models considered. In particular, the contribution of low-frequency modes where the dispersion relation is linear cannot be neglected in the system we are considering
(these photons can propagate in the crystal, being their frequency outside the gap): it actually causes the leading asymptotic power-law decay of the interaction energy.
Also, the possibility of controlling the entanglement of two qubits by coupling them to a common photonic bandgap environment has been recently investigated \cite{Wang-Jiang11}. Coupling quantum emitters to modes near the bandgap of a photonic crystal has allowed experimental investigation of the the strong coupling regime of quantum electrodynamics \cite{Liu-Houck17}, as well as observation of cooperative atom-atom interactions between emitters placed in a photonic crystal waveguide \cite{Hood-Goban16}.

\section*{The resonance interaction in a photonic bandgap environment}

The Hamiltonian of our system describes the interaction between a pair of two-level quantum emitters (atoms or quantum dots, for example) inside the photonic bandgap environment. In the multipolar coupling scheme and within the dipole approximation, it can be written as
\begin{eqnarray}
H &=& H_0 + H_I \, ,\nonumber \\
H_0 &=& H_A + H_B + \sum_{\bk j} \hbar \wk \akjd \akj  \, , \nonumber \\
H_I &=& - i \sum_{\bk j} \sqrt{\frac{2\pi \hbar \wk}\Omega} {\bf p}_A \cdot \ekj \! \left( \akj - \akjd  \right)
- i \sum_{\bk j}\sqrt{\frac{2\pi \hbar \wk} \Omega} {\bf p}_B \cdot \ekj \! \left( \akj e^{i\bk \cdot \br} - \akjd e^{-i\bk \cdot \br} \right) ,
\label{Hamiltonian}
\end{eqnarray}
where for the sake of generality we consider here both a three-dimensional and a one-dimensional system. A, B indicate the two atoms with positions $0$ and $\br$ respectively,
the polarization unit vectors $\ekj$ are assumed real, and $\Omega = V \, (L)$ in the three-dimensional (one-dimensional) case, with $V$ and $L$ being the quantization volume and length respectively.
$H_{A/B}=\hbar \wa \left( \mid e_{A/B}\rangle \langle e_{A/B} \mid - \mid g_{A/B}\rangle \langle g_{A/B} \mid \right)/2+\hbar \wa /2$ is the Hamiltonian of atom $A/B$, where $\mid e \rangle$ and $\mid g \rangle$ respectively indicate the atomic excited and ground state, and $\wa$ is the transition frequency of the atoms.

${\bf p}_A$ and ${\bf p}_B$ are the atomic operators which couple to the field: in three dimensions they are the atomic dipole moment operators
${\bm\mu}_{A\, (B)} = e\br_{A\, (B)}$,
while in the one dimensional case they have the dimension of a dipole moment per unit length and the wavevector $\bk$ and the positions $\br$ are both along the same direction. At the end we will take the continuum limit $\sum_\bk \rightarrow V/(2\pi )^3 \int dk k^2 d\Omega_\bk$ and
$\sum_\bk \rightarrow L/2\pi \int dk$, for the three- and one-dimensional cases, respectively.
In the dipolar coupling scheme \eqref{Hamiltonian}, static interatomic dipole-dipole interactions are already included in the interaction with the transverse field and thus mediated by photons \cite{Power-Zienau59}.
The presence of the external environment is fully taken into account through the specific photon dispersion relation and density of states.

In order to obtain the resonance interaction between the two identical emitters (atoms), we assume to prepare them in the following correlated symmetric or antisymmetric state
\begin{equation}
\mid \psi \rangle_{\pm} = \frac 1{\sqrt{2}} \left( \mid g_A, e_B; 0_{\bk j} \rangle \pm \mid e_A, g_B; 0_{\bk j} \rangle \right) ,
\label{state}
\end{equation}
where $g$ ($e$) indicates the ground (excited) state of the atom, and $0_{\bk j}$ is the photon vacuum.
The two states \eqref{state} are degenerate and perturbation theory for degenerate states should be used. However, from second-order energy corrections for degenerate states
\cite{Schiff68},
it is possible to see that the symmetric and antisymmetric subspaces do not mix with each other, provided the dipole matrix elements of the two atoms are equal and the photon dispersion relation is symmetric for $\bk$ and $-\bk$. If these conditions are met, the degenerate symmetric and antisymmetric states can then be treated separately and the second-order energy shift is given by non-degenerate perturbation theory in each subspace. These conditions are hold in our case, due to our dispersion relation and because ${\bf p}_A^{ge}={\bf p}_B^{ge}$,
where ${\bf p}_{A,B}^{ge}$ are the matrix elements of ${\bf p}_{A,B}$ between states $g$ and $e$.

The second-order energy shift is then given by
\begin{equation}
\sum_i \frac {\langle \psi^\pm \mid H_I \mid i \rangle \langle i \mid H_I \mid \psi^\pm \rangle}{E_{\psi^\pm}-E_i}
= \pm \Re \sum_i \frac {\langle g_A, e_B; 0_{\bk j} \mid H_I \mid i \rangle \langle i \mid H_I \mid e_A, g_B; 0_{\bk j} \rangle}{\hbar \wa-E_i}  + (L.S.) ,
\label{shift}
\end{equation}
where $\Re$ denotes the real part, $\mid i \rangle$ are intermediate states with energy $E_i$, $\hbar \omega_a = E_{\psi^\pm}$ is the energy of the atomic excited state,
and the $+$ or $-$ sign refers to the symmetric or antisymmetric state in \eqref{state}, respectively. $(L.S.)$ indicates energy corrections independent of the distance between the two atoms, i.e. the Lamb shift of the individual atoms. Eq. \eqref{shift} shows that the resonance interaction energy is strictly related to the resonant energy transfer between an excited and a ground-state atom, that is the excitation transfer between two atoms, whose probability amplitude is given by the absolute value of the first term in the RHS of \eqref{shift} \cite{Salam10,Weeraddana-Premaratne17}.
Therefore our results are also relevant for the resonant energy transfer between atoms.

Our purpose is to investigate the situation where the atomic transition frequency is inside the photon gap and in the proximity of its lower edge, so that the atom-field interaction can be safely described by second-order perturbation theory (the photonic gap reduces the resonant pole contribution). There are two possible intermediate states in Eq. \eqref{shift},
$\mid g_A, g_B; 1_{\bk j} \rangle , \, \, \, \mid e_A, e_B; 1_{\bk j} \rangle$,
which involve one real or virtual photon, respectively.

Since we can neglect distance-independent terms that do not contribute to the interatomic interaction, the second-order energy shift becomes
\begin{eqnarray}
\Delta E &=& \pm \Re \sum_{\bk j} \frac {2\pi \wk}\Omega (\ekj )_{\ell} (\ekj )_m ( p_A^{ge})_{\ell} ( p_B^{eg})_m
\left[ \frac {e^{i\bk \cdot \br}}{\wa -\wk} -  \frac {e^{-i\bk \cdot \br}}{\wa +\wk} \right] ,
\label{energyshift}
\end{eqnarray}
where $\wa$ is the atomic transition frequency and ${\bf p}_{A,B}^{ge}$ are matrix elements of the atomic operators between the ground and excited state, that we assume real.
Both rotating and counterrotating terms appear in Eq. \eqref{energyshift} as these terms give a comparable contribution in the low-frequency part of the integral.

An analysis has been carried out on both the three-dimensional and one-dimensional case. In the latter, it has become possible to experimentally obtain an environment with a photonic bandgap and a quadratic dispersion relation near the gap using nanophotonic waveguides, photonic crystals guides, coupled transmission line resonators or coupled cavities arrays \cite{Lodahl-Mahmoodian15}.
In the three-dimensional case, after the sum over polarizations, angular integration and in the continuum limit ($V \rightarrow \infty$), we obtain
\begin{equation}
\Delta E_{3D}= \pm \frac 1\pi ( \mu_A^{ge})_{\ell} ( \mu_B^{eg})_m \left( -\nabla^2 \delta_{\ell m} + \nabla_{\ell} \nabla_m \right) \frac 1r
\int_0^\infty \! \! dk \left[ \frac \wk{\wa -\wk}- \frac \wk{\wa +\wk} \right] \frac{\sin (kr)}k  ,
\label{energyshift3d}
\end{equation}
where $r$ is the distance between the atoms.

Similarly, in the one-dimensional case, after the sum over polarizations (the two polarizations are orthogonal to the distance between the atoms) and in the continuum limit ($L \rightarrow \infty$), the energy shift becomes
\begin{equation}
\label{energyshift1d}
\Delta E_{1D} = \pm 2  {\bf p}_{A \perp}^{ge}\cdot  {\bf p}_{B \perp}^{eg}
 \int_0^\infty dk \left[ \frac{\wk}{\wa -\wk}
-\frac{\wk}{\wa +\wk} \right]\cos (kr) ,
\end{equation}
where the subscript $\perp$ indicates the projection on the plane orthogonal to the interatomic distance.

Up to now, we have not yet specified the environment around the two atoms, except that $\wk$ depends only on $k=\mid \bk \mid$.
We now assume that the two atoms are embedded in an environment with a photonic band gap whose dispersion relation between frequency and wavevector is quadratic outside the gap and in the proximity of one of its edges, and linear for wavelengths much larger than the environment periodicity length. A quadratic dispersion relation is typical of a one-dimensional \cite{John-Wang90,Angelakis-Knight04,Kweon-Lawandy94} photonic crystal in the effective mass approximation, a coupled cavities array
\cite{Lodahl-Mahmoodian15,Hartmann-Brandao08}, or an impurity in a quantum wire near the gap \cite{Incardone-Fukuta14,Tanaka-Garmon06}.
It is commonly used to investigate radiative properties of atoms or quantum dots in a structured environment
\cite{ElGanainy-John13,Lodahl-Mahmoodian15,John-Quang94,Wang-Gu03}, for example a one-dimensional photonic crystal \cite{Goban-Hung14,Douglas-Habibian15}. In the isotropic three-dimensional case the same dispersion relation is assumed to be valid independently of the direction of propagation of the photon \cite{John-Wang91}.

We now consider the case of a one-dimensional (or, equivalently, isotropic three-dimensional) photonic crystal made by an infinite sequence of periodic nondispersive and nondissipative dielectric slabs with refractive index $n$ and thickness $2a$. They are separated by a distance $b$ of vacuum space, therefore the periodicity of the crystal is  $L=2a+b$. The crystal is periodic in one dimension and homogeneous in the two other directions. In such a case the dispersion relation can be obtained analytically when
$b = 2na$ in the following form \cite{John-Wang91,Angelakis-Knight04}
\begin{equation}
\wk = \frac c{4na} \arccos \left[  \frac {4n \cos (kL)+(1-n)^2}{(1+n)^2} \right] .
\label{disprel}
\end{equation}
The first gap occurs at $k_0= \pi /L$ with $L=2a(n+1)$ (the other gaps are at $k=m\pi /L$ with $m$ an odd integer number), and the lower and upper edges of the first gap are given by
\begin{equation}
\omega_\ell = \frac c{4na} \arccos \left( \frac {1+n^2-6n}{1+n^2+2n} \right)
\label{firstloweredge}
\end{equation}
\begin{equation}
\omega_u = \frac{\pi c}{2na}- \frac c{4na} \arccos \left( \frac {1+n^2-6n}{1+n^2+2n} \right)
\label{firstupperedge}
\end{equation}
Near the lower edge of the gap, the quadratic dispersion relation is
\begin{equation}
\wk = \omega_\ell - A(k-k_0)^2  ,
\label{dispersionrelation-l}
\end{equation}
while in the region above the upper edge of the gap is
\begin{equation}
\wk = \omega_u + A(k-k_0)^2  ,
\label{dispersionrelation-lII}
\end{equation}
where $A$ is a positive constant that depends
on the physical parameters of the environment (effective mass approximation). For a photonic crystal we get
\begin{equation}
A= \frac{(n+1)^2ac}{2(n-1)n^{1/2}} .
\label{constantA}
\end{equation}

For a photon with wavelength much larger than the periodicity of the crystal, $kL \ll 1$, the dispersion relation \eqref{disprel} gives the following linear relation
\begin{equation}
\wk \simeq \frac c{n^{1/2}} k ,
\label{dipersionrelationlow}
\end{equation}
where the factor $c/n^{1/2}$ is a sort of effective propagation velocity of low-frequency photons in the crystal: it takes into account the presence of the dielectric slabs of the photonic crystal which reduces the speed of light.

In order to be able to evaluate analytically the integrals over $k$ in Eqs. \eqref{energyshift3d} and \eqref{energyshift1d} in the range from $0$ to $k_0$, we interpolate the linear and quadratic behaviors of Eqs. \eqref{dipersionrelationlow} and \eqref{dispersionrelation-l} respectively, in the following form
\begin{equation}
\wk \simeq \alpha \frac c{n^{1/2}} k -\beta \frac{(n+1)^2ac}{2(n-1)n^{1/2}}k^2 ,
\label{interpolateddisprel}
\end{equation}
where $\alpha$ and $\beta$ are positive constants chosen appropriately to obtain a good approximation to the exact dispersion relation \eqref{disprel} in the complete range below the first gap ($0<k<k_0$). Requiring that the slope of the dispersion relation at $k=k_0=\pi /L = \pi /(2a(n+1))$ is zero as in the exact dispersion relation, we get
$\beta =2(n-1)\alpha/(\pi(n+1))$ and thus
\begin{equation}
\wk \simeq \frac {\alpha c}{n^{1/2}} \left( k-\frac {(n+1)a}\pi k^2\right) = \frac {\alpha c}{2n^{1/2}k_0} \left( 2kk_0-  k^2\right) ,
\label{interpolateddisprel1}
\end{equation}
where the value of the parameter $\alpha$ can be set to
$\alpha= (n+1)/(\pi \sqrt{n}) \arccos [(1+n^2-6n)/(1+n^2+2n)]$
in order to obtain the correct value of the frequency of the lower edge of the gap given by \eqref{firstloweredge}.
For $n=3$ we obtain $\alpha = 1.5396$.

\begin{figure}
\centering
\includegraphics[width=15cm]{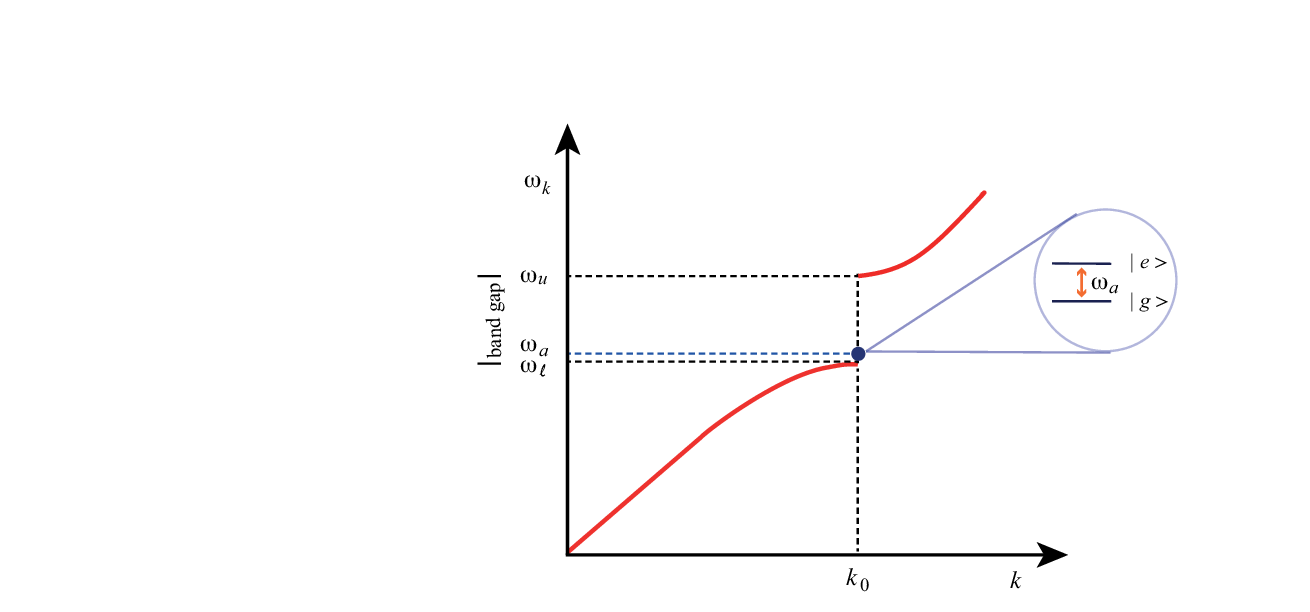}
\caption{The figure shows the position of the atomic frequency $\wa$, in relation to the lower and upper edge of the first photonic gap at wavenumber $k_0$, given by $\omega_\ell$ and $\omega_u$ respectively. We assume that the width of the photonic gap, $\Delta \omega = \omega_u - \omega_\ell$, is such that $\omega_a -\omega_\ell \ll \Delta \omega$.}
\label{Fig:1}
\end{figure}

Fig. \ref{Fig:1} shows the physical situation we are considering, where the atomic frequency $\omega_a$ is just above the lower edge $\omega_\ell$ of the gap at $k_0$,
and well below its upper edge $\omega_u$. The relative density of states can be obtained from the dispersion relation.
We wish to stress that the results we will obtain are not limited to the photonic crystal case, but valid for any bandgap environment with a dispersion relation linear at low frequencies and quadratic near the edge of the gap.

In the integrals over $k$ in \eqref{energyshift3d} and \eqref{energyshift1d}, all values of $k$ should in principle be included. However, only a much smaller range of $k$ gives a significant contribution in our case.
The density of states vanishes inside the gap (that is for photon frequencies between $\omega_\ell$ and $\omega_u$), while it diverges just outside both edges of the gap \cite{Lodahl-Mahmoodian15,John-Wang90}. If the atomic frequency $\omega_a$ is inside the gap, $\omega_\ell  <  \wa < \omega_u$, and very close to its lower edge $\omega_\ell$
($\wa - \omega_\ell \ll \Delta \omega = \omega_u - \omega_\ell$) as shown in Fig. \ref{Fig:1},
the modes above the gap give a quite smaller contribution to the integrals compared to the modes below the gap. This is due to both the near-resonance effect and the large density of states for the modes just below the gap. Moreover, low-frequency modes can give a relevant contribution at large distances between the atoms, exactly as in dispersion interactions \cite{Salam08,Compagno-Passante-Persico95}.
These arguments can be also verified by a direct comparison between the contribution of field modes with $0 \leq k \leq k_0$ (first Brillouin zone) and the contribution of modes in some frequency range above $k_0$. In our calculations we choose the range $k_0 \leq k \leq 3k_0/2$, where the upper extreme is half way between the first and the second gap of the photonic crystal, with the assumption of the quadratic dispersion relation \eqref{dispersionrelation-lII} above the first gap.
We now show that the latter contribution is indeed quite smaller than the former. This also indicates that we can safely neglect the contribution of the remaining modes with $k > 3k_0/2$. The rapid convergence of the integral over the higher Brillouin zones is also confirmed by the numerical calculation discussed later on in this section.

In the three-dimensional isotropic case, where the dispersion relation \eqref{interpolateddisprel1} is assumed valid independently of the direction of propagation and polarization of the photon, the $k$ integral in \eqref{energyshift3d} is given by
\begin{equation}
I_{3D} =  \int_0^{k_0} \! dk \left[ \frac{B(2kk_0-k^2)}{\wa -B(2kk_0-k^2)} -\frac{B(2kk_0-k^2)}{\wa +B(2kk_0-k^2)} \right] \frac {\sin (kr)}k ,
\label{equation-3d}
\end{equation}
while in the one-dimensional case the integral over $k$ in \eqref{energyshift1d} becomes
\begin{equation}
I_{1D} = \int_0^{k_0} \! dk \left[ \frac{B(2kk_0-k^2)}{\wa -B(2kk_0-k^2)}  -  \frac{B(2kk_0-k^2)}{\wa +B(2kk_0-k^2)} \right]  \cos (kr),
\label{equation-1d}
\end{equation}
where $B= \alpha c/(2n^{1/2}k_0)$ and only the range of $k$ from $0$ to $k_0$ has been considered in both expressions.

We now consider the large-distance limit $r \gg k_0^{-1} \simeq c/\wa$,
where the interaction energy has a sharp quantum nature.
This limit is indeed the most important from a physical point of view, as it gives an interatomic energy with a very long range (as $r^{-1}$ in free space \cite{Craig-Thirunamachandran98,Salam10}). It is related to the quantum coherence property of the state considered, while the short distance limit is essentially of electrostatic nature.
In this limit, the integrals over $k$ in \eqref{equation-3d} and \eqref{equation-1d} can be evaluated as
\begin{equation}
I_{3D} \simeq - \Gamma_3 \frac{\cos (k_0r)}{k_0r}  ,
\label{equation1-3}
\end{equation}
where
\begin{equation}
\Gamma_3 = \frac {2\wa B k_0^2}{\wa^2-B^2k_0^4}+\sqrt{\frac {Bk_0^2}{\wa+Bk_0^2}}>0
\label{gamma3}
\end{equation}
is a dimensionless constant expressed in terms of the physical parameters of the system, and
\begin{equation}
I_{1D} \simeq \Gamma_1 \frac{\sin (k_0r)}r  ,
\label{equation1-1}
\end{equation}
where we have introduced the dimensionless constant
\begin{equation}
\Gamma_1 = 2 \left( \frac {\wa B k_0^2}{\wa^2-B^2k_0^4}+1 \right)>0 .
\label{gamma1}
\end{equation}

We can now compare the contributions obtained before with those coming from the integration above the (first) gap, for example between the first and the second gap in the case of a photonic crystal. In this range, we expect that values of $k$ closer to $k_0$ should give a larger contribution with respect to values near the next gap, as they are closer to resonance.
Hence we may evaluate the contribution of the range of $k$ from $k_0$ to $3k_0/2$,
using the dispersion relation in the effective mass approximation above the gap \eqref{dispersionrelation-lII}: the upper limit could be anywhere within the second band without significantly changing our results. The relative $k$ integrals can be evaluated analytically in terms of the system parameters $\wa ,k_0,B,\omega_\ell ,\omega_u$: it is possible to use typical values for the transition frequency in the optical region and a typical photonic crystal for the environment.

The use of typical values such as $n=3$ and $a=2 \cdot 10^{-8} \text{m}$ in Eqs. \eqref{firstloweredge} and \eqref{firstupperedge} gives
$\omega_\ell = 2.616 \cdot 10^{15} \text{s}^{-1}$ and $\omega_u = 5.232 \cdot 10^{15} \text{s}^{-1}$ for the frequency of the lower and upper edge respectively, which are found in the optical region of the spectrum. Choosing the atomic transition frequency $\omega_a = 2.65 \cdot 10^{15} \text{s}^{-1}$, that is just above the lower edge of the gap and in the optical region, the width of the photonic gap
$\Delta \omega = \omega_u-\omega_\ell$ is such that $\omega_a -\omega_\ell \ll \Delta \omega$, as we have assumed. Also, we have in this case $\alpha = 1.5396$ and the quantity $\frac {\alpha c}{n^{1/2}}$ in Eq. \eqref{interpolateddisprel1}, which gives the effective propagation speed for low-frequency photons in the crystal, has the value $\simeq 0.89 c$.
With these numerical values the contribution from the integration range $0 \leftrightarrow k_0$ in the three-dimensional case is several orders of magnitude larger than the corresponding contribution from the integration range $k_0 \leftrightarrow 3k_0/2$,
while in the one-dimensional case it is about two orders of magnitude larger, thus allowing the latter contribution to be neglected in both cases.
The closer the atomic frequency is to the lower edge of the gap, the better the approximation of neglecting contributions with $k>k_0$.
The previous discussion also gives a strong indication that all contributions coming from $k>3k_0/2$ will be negligible, since they are more distant from resonance.
Moreover, they involve higher field frequencies, yielding a much smaller contribution of the related (virtual) transitions at large distance, due to the higher energy unbalance for the intermediate state.
This is also confirmed by a numerical calculation of the integrals over $k$ in the lowest Brillouin zones using the exact dispersion relation \eqref{disprel}, as we will discuss in the following.

Substitution of \eqref{equation-3d} and \eqref{equation1-3} into \eqref{energyshift3d} finally gives
\begin{equation}
\Delta E_{3D} \simeq \mp \frac {\Gamma_3}{\pi} k_0^3 ( \mu_A^{ge})_{\ell} ( \mu_B^{eg})_m \left( \delta_{\ell m} - \hat{r}_{\ell} \hat{r}_m \right) \frac{\cos (k_0r)}{(k_0r)^2} \, ,
\label{energyshift3}
\end{equation}
showing that the resonance interaction asymptotically scales as $r^{-2}$ inside our bandgap environment.

The result \eqref{energyshift3} should be compared to the three-dimensional free-space case\cite{Craig-Thirunamachandran98,Salam10}
\begin{equation}
\Delta E_{3D}^{f.s.} \simeq \mp \left( \frac {\wa} c \right)^3  ( \mu_A^{ge})_{\ell} (\mu_B^{eg})_m \left( \delta_{\ell m} - \hat{r}_{\ell} \hat{r}_m \right) \frac{\cos \left( \frac {\wa r}c \right)}{\frac {\wa r}c} ,
\label{energyshift3Dfs}
\end{equation}
where the resonant energy scales as $r^{-1}$ at large distances.

For the one-dimensional case, substitution of \eqref{equation1-1} and \eqref{equation-1d} into \eqref{energyshift1d} yields, in the long-distance limit,
\begin{equation}
\Delta E_{1D} \simeq \pm 2 \Gamma_1 {\bf p}_{A \perp}^{ge}\cdot  {\bf p}_{B \perp}^{eg} k_0 \frac {\sin (k_0r)}{k_0r}  .
\label{energyshift1}
\end{equation}

The result above should be compared with the analogous quantity obtained for atoms in the (one-dimensional) free space \cite{ElGanainy-John13,Incardone-Fukuta14}
\begin{equation}
\Delta E_{1D}^{f.s.} \simeq \pm 2\pi {\bf p}_{A \perp}^{ge}\cdot  {\bf p}_{B \perp}^{eg} \frac {\wa}c \sin \left( \frac {\wa r}c  \right)  .
\label{energyshift1Dfs}
\end{equation}

Our findings \eqref{energyshift3} and \eqref{energyshift1} should be also compared to the results obtained in Refs. \cite{Incardone-Fukuta14,Bay-Lambropoulos97} when the atomic transition frequency is outside the gap and in the proximity of one of its edges. In this case a strong enhancement of the interaction energy is obtained, due to the large density of states near the band edge (van Hove singularity), while the scaling with the distance is the same as in free space: $(k_0 r)^{-1}\cos (k_0 r)$ and $\sin (k_0 R)$ for the three-dimensional and one-dimensional cases respectively. In Ref. \cite{Bay-Lambropoulos97} the resonant dipole-dipole interaction has been also considered in the case of atomic frequency in the gap, just below its upper edge (on the contrary, our results refer to the case of the atomic frequency just above the lower edge of the gap).

These outcomes clearly show that the bandgap environment can qualitatively change the scaling of the interaction with the distance when the atomic transition frequency is within the photonic gap. The change of the scaling of the interaction in comparison with the free-space case comes from the suppression of the contribution of the resonance pole in the frequency integration: in fact the interaction is basically mediated by nonresonant photons.
In a quasi-static approach, the force between the two atoms can then be obtained by taking the opposite of the derivative of the energy shift \eqref{energyshift3} or \eqref{energyshift1} with respect to the interatomic distance.

Equations \eqref{energyshift3}  and \eqref{energyshift1} contain our main results.
Eq. \eqref{energyshift3} exhibits a large-distance $r \gg c/\wa \simeq k_0^{-1}$ scaling of the resonance energy as $r^{-2}$ in the isotropic three-dimensional case, with an oscillation which gives a spatially periodic change of the resonance force from attractive to repulsive, analogously to the one-dimensional case. Since in 3D free space the asymptotic behavior of the interaction energy given by Eq. \eqref{energyshift3Dfs} decays as $r^{-1}$, our result in Eq. \eqref{energyshift3} clearly reveals that the presence of the bandgap environment can qualitatively change the scaling of the interaction with the distance.
Analogously, Eq. \eqref{energyshift1} shows that in the one-dimensional case, for $r \gg c/\wa \simeq k_0^{-1}$, the resonant interaction scales with the interatomic distance as $r^{-1}$, again presenting a spatial oscillation which changes the resonance force from attractive to repulsive. Because in 1D free space the respective behavior is a pure oscillation (see Eq. \eqref{energyshift1Dfs}), our results indicate that also in the one-dimensional case the bandgap environment can qualitatively modify the dependence of the interaction energy on the distance.
\begin{figure}
\centering
\includegraphics[width=12cm]{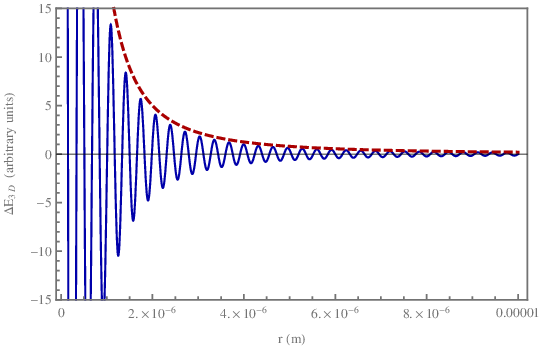}
\caption{Plot of the resonance interaction energy $\Delta E_{3D}$ in arbitrary units for distances of the atoms between $10^{-7} \text{m}$ and $10^{-5} \text{m}$ obtained by a numerical integration using the exact dispersion relation of the photonic crystal (blue continuous curve). The atomic dipoles are assumed equal and oriented perpendicularly to the atomic distance. Parameters are such that $n=3$, $a=2 \cdot 10^{-8} \text{m}$, $\omega_a = 2.65 \cdot 10^{15} \text{s}^{-1}$, and thus $k_0 = 1.96 \cdot 10^{7} \text{m}^{-1}$. The red dashed line represents the function $r^{-2}$. The figure clearly shows the asymptotic power-law decay as $r^{-2}$, in agreement with the analytical results obtained using our approximate dispersion relation in the first Brillouin zone of the photonic crystal.}
\label{Fig:2}
\end{figure}

\begin{figure}
\centering
\includegraphics[width=12cm]{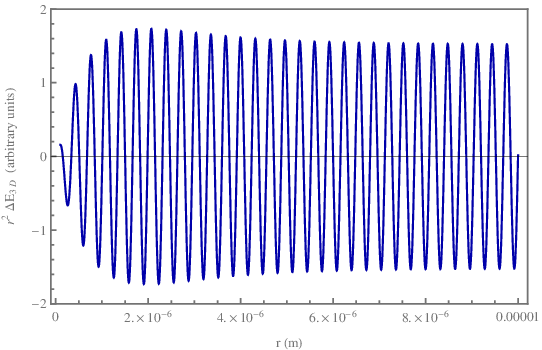}
\caption{Plot of $r^2\Delta E_{3D}$ in arbitrary units for distances of the atoms between $10^{-7} \text{m}$ and $10^{-5} \text{m}$ obtained by a numerical integration using the exact dispersion relation of the photonic crystal and the same parameters of Figure \ref{Fig:2}. The figure clearly shows that, after the transition region between the near and far zone at $r \sim 5 \cdot 10^{-6} \text{m}$, the function $r^2\Delta E_{3D}$ asymptotically settles to a constant value. This confirms that asymptotically the resonance interaction energy $\Delta E_{3D}$ scales as $r^{-2}$.}
\label{Fig:3}
\end{figure}

We have also checked numerically our power-law result by using the exact dispersion relation \eqref{disprel} for both one- and three-dimensional cases. In particular we performed a numerical evaluation of the integrals over $k$ resulting from Eq. \ref{energyshift}. Here we explicitly show the numerical result for the three-dimensional case, when identical dipoles are oriented perpendicularly to the interatomic distance (similar considerations apply in other configurations and for the one-dimensional case). In this case, Eq. \ref{energyshift} yields
\begin{equation}
\label{3Dnum}
\Delta E_{3D} = \pm \frac 1\pi \mid {\bm{\mu}}^{ge} \mid^2 \int_0^\infty \! dk \left( \frac \wk{\wa -\wk}- \frac \wk{\wa +\wk}\right)
\left( \frac {\cos (kr)}{r^2}- \frac {\sin (kr)}{kr^3}+ \frac {k \sin (kr)}{r} \right) .
\end{equation}

We have evaluated numerically the integrals over $k$ in \eqref{3Dnum} with the exact dispersion relation \eqref{disprel}, using the same values of the parameters chosen for our previous considerations on our analytical results. The periodicity of the photonic crystal is thus $L=1.6 \cdot 10^{-7} \text{m}$ and the first gap is at $k_0 = 1.96 \cdot 10^{7} \text{m}^{-1}$.
Figures \ref{Fig:2} and \ref{Fig:3} show in arbitrary units the outcome of the numerical calculation for distances $r$ from $10^{-7} \text{m}$ to $10^{-5} \text{m}$ for $\Delta E_{3D}$ and  $r^2\Delta E_{3D}$, respectively. This distance range goes from the transition region between near and far zone to the full far zone. Figure \ref{Fig:2} shows the oscillations of the interaction energy: in particular its envelope in the far zone coincides with the red dashed curve, which represents the function $r^{-2}$. This makes clear the asymptotic behaviour of $\Delta E_{3D}$ as $1/r^2$, in full agreement with our power-law analytical result \eqref{energyshift3}. Moreover, Figure \ref{Fig:3} explicitly shows that $r^2\Delta E_{3D}$ settles to a constant value after the transition zone between near and far zone at $r \sim k_0^{-1}$. Our numerical evaluation of \eqref{3Dnum} in the first and next Brillouin zones has also allowed us to check the rapid convergence of the $k$ integral over the Brillouin zones, in agreement with our previous analytical considerations.

To summarize, we have considered the case where the atomic transition frequency is within the band gap of the structured environment: in this configuration the resonance interaction decreases more rapidly with the distance compared to the free space case. Despite that, the important aspect to stress is that the spontaneous decay of the atoms is strongly suppressed by the presence of the environment (see also \cite{Lodahl-vanDriel04,Wang-Stobbe11}): this is crucial to preserve the quantum coherence of the correlated state from the decoherence effects due to spontaneous emission. For example, in Refs. \cite{Noda-Fujita07,Jorgensen-Galusha11} a suppression of the spontaneous decay rate of more than a factor $10$ in the optical region has been obtained by using a photonic crystal.
Also, inhibition of spontaneous emission of quantum dots in the near infrared region by two or more orders of magnitude has been obtained in recent experiments \cite{Wang-Stobbe11} when they are placed in photonic nanostructures. In our case this would guarantee the ability to significantly extend the lifetime of the correlated state of the two atoms.

Our findings display a large-distance scaling of the interaction energies \eqref{energyshift3} (three-dimensional case) and \eqref{energyshift1} (one-dimensional case) with the interatomic distance as a power law. This result should be compared with other recent results in the literature for the dipole-dipole interaction in a photonic crystal waveguide, where an exponential behavior of the related dipole-dipole interaction is obtained \cite{Shahmoon-Kurizki13,Douglas-Habibian15,Shahmoon-Grisins16,John-Wang91}. The main difference between these systems and the one under our consideration is that in our case the photonic crystal is homogeneous in the two directions orthogonal to that with the periodical dielectric slabs, and has a very large extension (in particular compared to the atomic transition wavelength $c/\wa$). This is a crucial point as implies that there is not a lower cut-off frequency (see discussion around Eq. \eqref{interpolateddisprel1}), therefore
our environment is quite different from a (photonic crystal) waveguide. Under these conditions, a relevant contribution to the interaction energy comes from low-frequency modes: they  give a power-law behavior even if virtual transitions are involved, since the resonant transition is suppressed by the presence of the photonic gap. This is similar to what happens in dispersion forces between ground-state atoms \cite{Compagno-Passante-Persico95,Craig-Thirunamachandran98}. In our case the fact that a single low-frequency mode has a factor $1/V$, with $V$ the quantization volume \cite{Kurizki-Genack88}, does not reduce the role of these modes because the integration over all continuous modes below the first gap at $k=k_0$ introduces a multiplicative factor $V$.
Some previous results in the literature indicate an exponential decay of the dipole-dipole resonance interaction \cite{John-Wang91} for the same model of photonic crystal as ours. However, the analytical result reported in Ref. \cite{John-Wang91} is based on the effective mass approximation and a quadratic dispersion relation yielding a nonvanishing frequency at $k=0$ was used for all frequencies considered. Such a dispersion relation does not take into account properly the contribution of  photons with $kL \ll 1$ though, which have a linear dispersion relation, as we have shown above. These photons give an important contribution to the resonance interaction at large distances and their linear dispersion relation is at the origin of the power-law scaling we find for large interatomic distance.
The numerical results in Ref. \cite{John-Wang91} cannot be immediately compared to our power-law results, because there the plots of the dipole-dipole interaction are given in a small range of distances. In addition the typical analytical asymptotic behavior given in \cite{John-Wang91} contains the localization length, that is related to resonant photons bound to the atom because they cannot propagate in the photonic crystal. This indicates that probably only the contribution of the guided modes close to the atomic resonance has been included.
Similar considerations hold also for the strong suppression of the interaction inside the gap found in Ref. \cite{Kurizki-Genack88} (see also the comment in Ref. \cite{Zheng-Kobayashi96} and the reply in Ref. \cite{Kurizki-Kofman96}).
Our results for the resonance interaction energy have been obtained for identical atoms, assuming the symmetrical or antisymmetrical state given in Eq. \eqref{state}. Our procedure requires that the transition frequency of the two atoms be nearly equal and both close to the lower band edge of the photonic crystal. The (effective) coupling strenght with the field can however be different for the two atoms; this situation could occur even for identical atoms in the case of spatial inhomogeneities of the environment yielding different environment's local properties at the atomic positions. This situation could be described through an effective value of the coupling constant and, in such a case, our results \eqref{energyshift3} and \eqref{energyshift1} are still valid, using the appropriate values of the effective coupling constants for the two atoms.

We can conclude that a possible experiment aiming to observe the resonance interaction energy and force (not directly observed yet) could be performed including such environment. Although its presence reduces the interaction for $\wa r/c \gg 1$, this effect is sensibly counteracted by the much stronger stability of the correlated state provided by the inhibition of the spontaneous decay induced by the environment: this effect allows to observe the system for a much longer time. This could be an essential point in designing an experiment for a direct observation of the resonance force, strongly suggesting that a photonic bandagap environment can provide a suitable setup when the atomic frequency is inside the bandgap and close to {its lower edge. Furthermore, using the typical values of the system parameters given after Eq. \eqref{gamma3}, our results show that the bandgap environment increases the interaction for distances up to $\sim 25 c/\wa$, although it decreases with respect to the free-space case at larger distances.
A critical point for an experimental realization of the setup we are proposing rests on how to prepare the (symmetric or antisymmetric) entangled state inside the photonic crystal, given that photons resonant with the atomic frequency cannot propagate inside the crystal. A possibility could be to send photons with a frequency lower than the atomic one, which can propagate in the photonic crystal as the bandgap does not affect them. Then a multi-photon absorption by the atoms could be exploited to excite one of them and create the correlated state. Trapping of cesium atoms in a photonic crystal waveguide by tight optical potentials has been recently achieved to observe superradiance from entangled atoms and strong coupling \cite{Goban-Hung15}.

\section*{Discussion}

In this paper we have considered the resonance interaction energy between two entangled atoms, one excited and the other in its ground state, prepared in their symmetric or antisymmetric state and interacting with the electromagnetic field in the vacuum state. The two atoms are placed in a generic photonic bandgap environment with a quadratic dispersion relation and their transition frequencies inside the gap and in the proximity of its lower edge. Several environments, such as photonic crystals, an array of coupled optical cavities, nanophotonic waveguides and coupled transmission line resonators, are common experimental realizations of the environment we are considering.  We have stressed the formal similarity of the expression for the interaction energy with the excitation transfer between atoms, which provides a direct extension of our results to this problem too. In our case the spontaneous decay of the atoms is strongly suppressed due to the small photon density of states inside the gap. We have shown that the presence of the environment changes the asymptotic distance dependence of the resonance interaction energy (where quantum coherent effects are essential) from $1/r$ to $1/r^2$ in an isotropic three-dimensional case, and from a pure oscillation to $1/r$ in the one-dimensional case.
We have also checked our analytical results obtained with an approximated dispersion relation with a numerical calculation using the exact dispersion relation, obtaining a full agreement between the two approaches.
We compare our outcomes with previous results in the literature for the dipole-dipole interaction of atoms in photonic crystals and photonic crystal waveguides and highlight the differences on the scaling of the interaction energy with the distance. Our results
explicitly show the possibility to control the resonance interaction energy and force using the bandgap environment. In the case considered in this paper the resonance force between the two atoms can be reduced with respect to the free space case for very large distances. However, it is still much stronger than other long-range interactions, such as dispersion interactions, even when factorized excited states are considered. More importantly, the most meaningful point here is that the decoherence effect of the spontaneous decay is significantly reduced. Strong inhibition of the spontaneous emission rate of about two orders of magnitude has been recently measured due to the presence of structured environments such as photonic crystals \cite{Wang-Stobbe11}. Thus the quantum coherent atomic superposition can live for an extended period of time thanks to the presence of the environment. Its lifetime is of the same order of the excited state lifetime in the environment considered, which allows possible detection of this elusive coherent quantum effect. Therefore our findings are relevant for designing a future experiment aiming to directly measure the resonance force, which has not been observed yet, using atoms or quantum dots embedded in an appropriate structured environment with a photonic bandgap.


\section*{Acknowledgements}

The authors gratefully acknowledge financial support by the Julian Schwinger Foundation and by MIUR. The authors wish to thank Richard Stones for his careful reading of the manuscript.

\section*{Author contributions statement}

R.P. and L.R. conceived the original idea of the problem considered and supervised the research project. V.N. performed most of the calculations and developed the main theoretical analysis of the system. R.P. wrote the manuscript. All authors equally participated to develop the research project and to discussions on the results obtained and their interpretation. All authors contributed to review and revise the manuscript.

\section*{Additional information}

\textbf{Competing interests} The authors declare no competing interests.



\end{document}